\documentclass[column,prb,floats,showpacs,superscriptaddress]{revtex4}
\usepackage{graphicx,epsfig}
\usepackage{times}
\usepackage{graphics,dcolumn,bm,float}
\usepackage{amssymb,amsmath,rotate,color}

\begin{document}
\unitlength 1 cm
\newcommand{\be}{\begin{equation}}
\newcommand{\ee}{\end{equation}}
\newcommand{\bearr}{\begin{eqnarray}}
\newcommand{\eearr}{\end{eqnarray}}
\newcommand{\nn}{\nonumber}
\newcommand{\vk}{\vec k}
\newcommand{\vp}{\vec p}
\newcommand{\vq}{\vec q}
\newcommand{\vkp}{\vec {k'}}
\newcommand{\vpp}{\vec {p'}}
\newcommand{\vqp}{\vec {q'}}
\newcommand{\bk}{{\bf k}}
\newcommand{\bp}{{\bf p}}
\newcommand{\bq}{{\bf q}}
\newcommand{\br}{{\bf r}}
\newcommand{\bR}{{\bf R}}
\newcommand{\up}{\uparrow}
\newcommand{\down}{\downarrow}
\newcommand{\fns}{\footnotesize}
\newcommand{\ns}{\normalsize}
\newcommand{\cdag}{c^{\dagger}}

\title{Giant Magnetoresistance in Bilayer Graphene Nanoflakes}
\author{Rouhollah Farghadan}\email{rfarghadan@kashanu.ac.ir}
\affiliation{Department of Physics, University of Kashan, Kashan, Iran }
\author{Marzieh Farekiyan}
\affiliation{Department of Physics, University of Kashan, Kashan, Iran }
\date{\today}

\begin{abstract}
Coherent spin transport through bilayer graphene (BLG) nanoflakes sandwiched between two electrodes made of single-layer zigzag graphene nanoribbon was investigated by means of Landauer-Buttiker formalism. Application of a magnetic field only on BLG structure as a channel produces a perfect spin polarization in a large energy region. Moreover, the conductance could be strongly modulated by magnetization of the zigzag edge of AB-stacked BLG, and the junction, entirely made of carbon, produces a giant magnetoresistance (GMR) up to $10^6\%$. Intestinally, GMR and spin polarization could be tuned by varying BLG width and length. Generally,  MR in a AB-stacked BLG strongly increases (decreases) with length (width).
\end{abstract}

\maketitle
\section{Introduction}
Graphene systems, which are composed of one or a few carbon monatomic layers, have attracted much attention due to their unusual electronic properties even at room temperature, such as a high carrier mobility, thermal and structural stabilities, low spin orbit, hyperfine interactions, long spin relaxation length, and gate-tunable spin transport, and their potential for applications in nanoelectronic and spintronic devices \cite{Oostinga,Ghosh,Nguyen,Echtermeyer,Saffarzadeh}.Currently, there is a continuous interest in BLG properties, both theoretically and experimentally. A bilayer graphene consists of two single graphene layers and has two different stacking arrangements: AB (Bernal) and AA. Energy band gap  in BLG nanoribbon and nanoflake could be controlled by substrate properties and the applied vertical electric field, so these structures could be used as a channel material in carbon-based transistors \cite{Zhong,Li,Xia,Yang,Han,Sato,McCann}.

Moreover, several studies have been performed, both theoretically and experimentally, on transport properties in BLG structure \cite{Zhong,Li,Xia,Yang,McCann,Gregersen,San-Jose,Ouyang,Pereira,Zhang1}. Interestingly, BLG field-effect-transistors with high on/off current ratios at room temperature have been reported \cite{Yang}. In BLG nanoribbons with zigzag edges, electronic transport is dominated by edge-states similar to those of single-layer graphene  \cite{Castro,Pisani,Sahu1,Sahu2,Lee,Guo}.

These states are expected to be spin-polarized and make zigzag-edge BLG nanoribbons' junctions attractive for spin-polarized transport \cite{Sahu1,Sahu2,Lee,Guo}. On the other hand, long spin relaxation length in bilayer graphene nanoribbons and few-layer graphene flakes was observed at room temperature \cite{Gui}. Moreover, spin-filtering and spin-dependent transport properties of BLG with ferromagnetic electrode and gate, external magnetic field, and exchange field have been investigated theoretically and experimentally \cite{Ghosh,Nguyen,Yang,Han,Orellana,Yu,Liu}. Another important feature of graphene in spintronic devices is its magnetoresistance. Giant magnetoresistance (GMR) in graphene junction with ferromagnetic electrodes \cite{Bai,Zhang,Gopinadhan,Farghadane} and magnetic states of zigzag edges have been extensively investigated \cite{Kim,Munoz-Rojas,Li2015,Wang}. But there has been a little attention to the magnetoresistance effects in BLG junction \cite{Nguyen}.

In this paper, we studied the coherent magnetic transport properties of a BLG nanoflake connected to two single-layer zigzag graphene nanoribbon (ZGNR) electrodes. The results showed that the application of a magnetic field only on BLG as a channel changes the spin configuration and conductivity of the BLG zigzag edge and induces a perfect spin-polarized conductance in the AB- stacked BLG structures. Moreover, we studied the magnetic transport properties by changing BLG width and length connected between ZGNR electrodes. Finally, we found a GMR up to $10^6\%$ and suggested that the GMR could be tuned by varying BLG width and length.

\section{MODEL AND METHOD}

We simulated the system depicted in Fig. 1a using a $\pi$-orbital tight-binding model and Hubbard repulsion treated in the mean-field approximation. This formalism that includes $e-e$ interaction in BLG induces localized magnetic moments in zigzag-edge atoms. We wrote the mean-field Hamiltonian in the AB-stacked BLG as follows \cite{Saffarzadeh}:

\begin{equation}
H_{C}= t\sum_{<i,j>,\sigma} c^{\dagger}_{i,\sigma}c_{j,\sigma}
+U\sum_{i,\sigma}\hat{n}_{i,\sigma}\langle \hat{n}_{i,-\sigma}\rangle \,
\end{equation}

where $c^{\dagger}_{i\sigma}$ ($c_{i\sigma}$) stands for the creation (annihilation) operator of an electron with spin $\sigma$ located on site. The tight-binding parameters $t$ were fixed to their bulk values equal to $t=-2.66 eV$ for the in-plane nearest neighbors of i and j and $t=-0.35 eV$ for the interlayer hopping parameter  \cite{Nilsson}, and $U=2.82 eV$ is the on-site coulomb interaction. The Green�s function of the junction is expressed as

\begin{equation}
\hat{G}_{C,\sigma}(\varepsilon)=[\varepsilon\hat{I}-\hat{H}_{C,\sigma}-\hat\Sigma_{S}(\varepsilon)-\hat\Sigma_{D}(\varepsilon)]^{-1}.
\end{equation}
where $\hat\Sigma_{S}(\varepsilon)$, $\hat\Sigma_{D}(\varepsilon)$ are the self-energy matrices due to the connection of right and left ZGNR electrodes to the channel, respectively. The spin-dependent
density of states (DOS) of the BLG in the presence of electrodes is given by

\begin{equation}
D_\sigma(\varepsilon)=-1/\pi Im \langle{\sigma |G_{C,\sigma}(\varepsilon)|\sigma}\rangle.
\end{equation}

Accordingly, the magnetic moment at each atomic site can be expressed as
$m_i= \mu_B (\langle \hat{n}_{i,\sigma}\rangle-\langle \hat{n}_{i,-\sigma}\rangle)$
 According to the Landauer-Buttiker formalism \cite{Buttiker,Farghadane}  the spin-dependent conductance and currents can be written as
\begin{equation}
G^{\sigma}(\varepsilon)= (e^2/h) {Tr}[\hat{\Gamma}_{S}
\hat{G}_{C,\sigma}\hat{\Gamma}_{D}\hat{G}_{C,\sigma}^{\dagger}].
\end{equation}
 Using $\hat\Sigma_{(\varepsilon)}$, the coupling matrices $\hat{\Gamma}_{\alpha}$ can  be expressed as $\hat{\Gamma}_{\alpha}=-2 Im[\hat\Sigma_{(\varepsilon, \alpha)}]$. Also, the spin-dependent currents can be written as

\begin{equation}
I_\sigma=1/e \int_{-\infty}^{\infty} {G_{\sigma}(\varepsilon)(f_{S}(\varepsilon-\mu_L)-f_{D}(\varepsilon-\mu_R))d\varepsilon. }
\end{equation}
where $f$ is the Fermi$�$Dirac distribution function and consider the effect of temperature.

We calculate the magnetoresistance of each device as $
MR= (I_{FM}-I_{AFM})/I_{AFM} \times 100.$

We solved equation 1-3 self consistently by an iteration method. By choosing anti-ferromagnetic (AFM) and ferromagnetic (FM) spin configurations as initial conditions in BLG nanoflake, different spin configurations of AFM and FM were achieved, respectively. Moreover, the effect of single-layer graphene nanoribbon as an electrode on BLG was added via self-energies, and the Green function of BLG was subsequently calculated. Finally, the new expectation values of number operators were replaced in the Hamiltonian, and this process was repeated until convergence was achieved\cite{Farghadan1}.
Note that in the bilayer graphene nanoflake, the upper single-layer graphene was only connected to the lower one through single-band tight binding model in the first nearest neighbors' approximation. The effect  of electrodes on the upper monoatomic sheet was indirectly added through the hopping parameter between two monatomic sheets in BLG nanoflake. Modulation of hopping parameter between electrodes and BLG could broaden electron energy level and consequently lead to the decrease of edge magnetism in BLG nanoflakes\cite {nanotechnology}.

\section{RESULTS AND DISCUSSION }

\begin{figure}
\centerline{\includegraphics[width=0.9\linewidth]{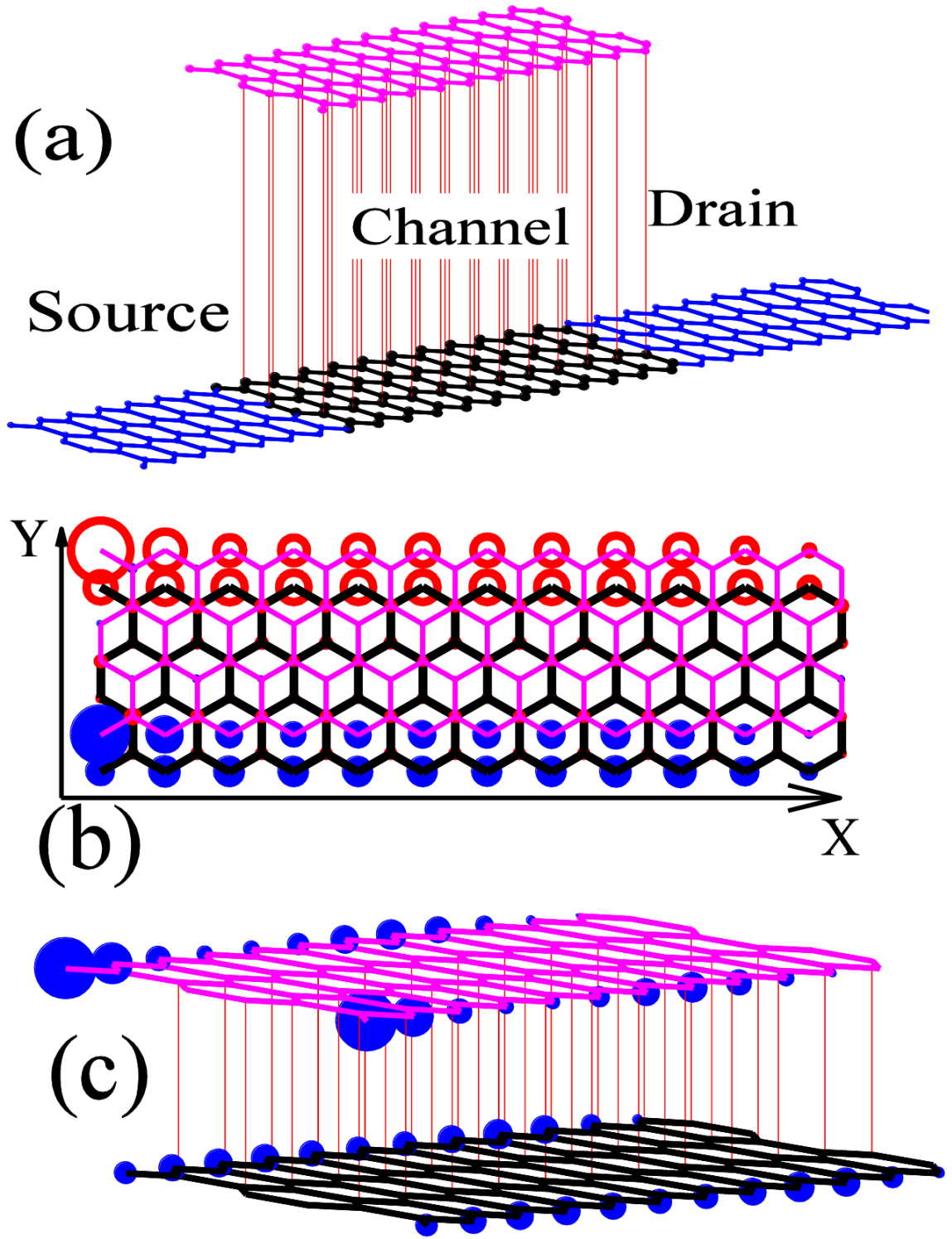}}
\caption{(a) Schematic view of BLG nonoflake with the width $N_y=4$ and length $N_x=12$ attached between semi-infinite ZGNR electrodes, (b) (top view) and (c)  AB-stacked BLG as a channel with AFM and FM spin configurations on zigzag edges, respectively. Open (red in online version) and filled (blue in online version) circles represent spin-down and spin-up densities, respectively. The magnetic moments are within [-0.25:0.25] $\mu_B$ and [-0.05:0.25] $\mu_B$ in cases b and c , respectively.
}
\end{figure}

\begin{figure}
\centerline{\includegraphics[width=1\linewidth]{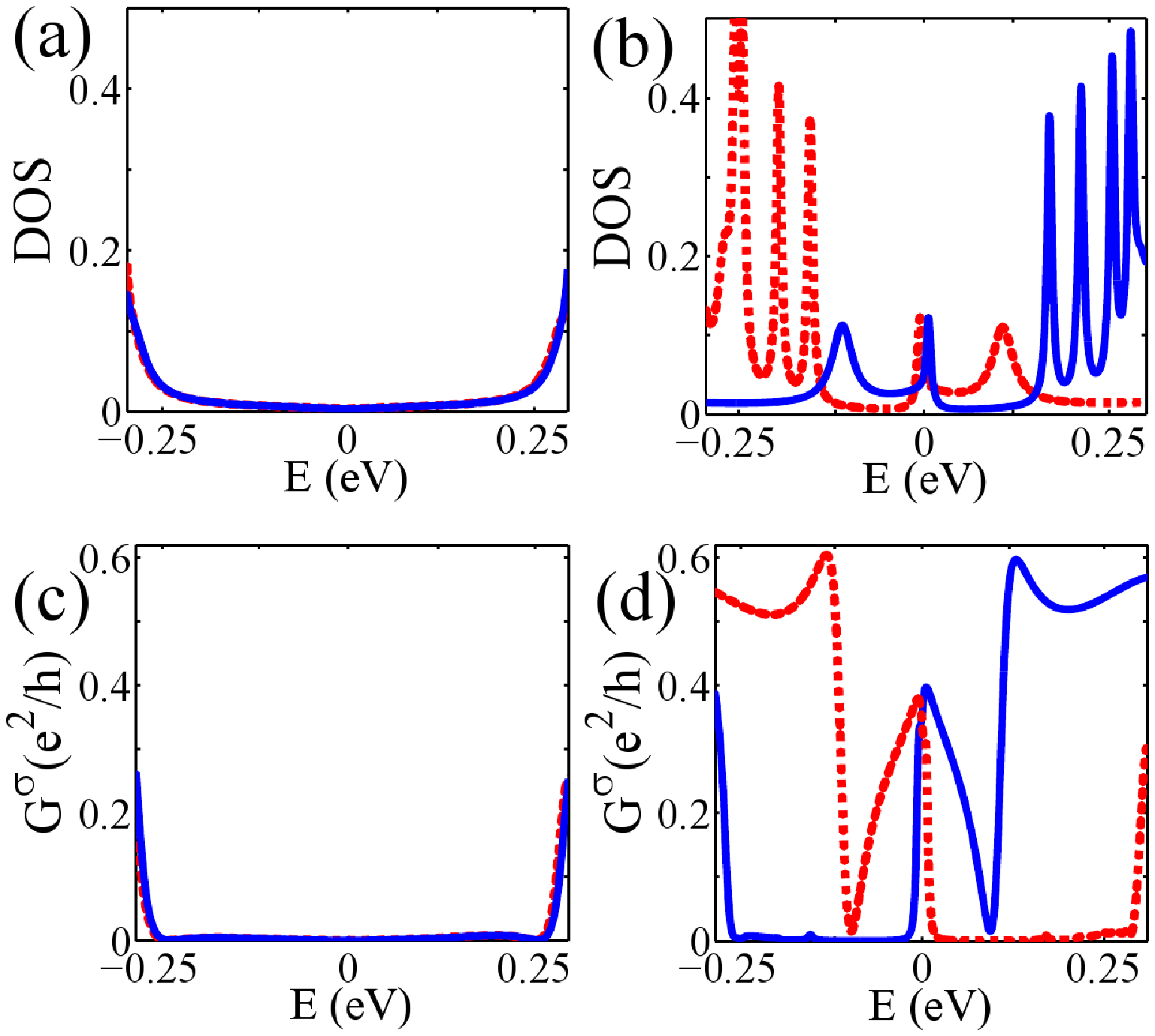}}
\caption{The electronic and magnetic transport properties of the ZGNR/ BLG/ZGNR junction depicted in Fig. 1 ($N_y=4$ and $N_x=12$) with AFM (a and c) and FM (b and d) spin configurations. (a) and (b) show the density of states, and (c) and (d) show the transmission coefficient of the BLG junction. The solid (dashed) line is for minority- (or majority-) spin electrons. }
\end{figure}

We considered AB-stacked BLG as a channel connected between two semi-infinite single-layer ZGNR electrodes. A similar junction (monolayer-bilayer junction) was proposed in previous works and its electronic and transport properties were investigated even under applied biases  \cite{Habib, Nilsson,Nakanishi,Koshino,Gonzalez,Yin}. The number of armchair rows along x direction, $N_x$, defines a BLG length and the number of zigzag rows along y direction, $N_y$, defines a BLG width. Thus, BLG length and width are $L=N_xa$ and  $W=(\sqrt{3}/2)N_ya$, respectively, with a=2.42\AA (Fig. 1b).

Anti-ferromagnetic (AFM) spin configuration of AB-stacked BLG (see Fig. 1b) is more stable than other spin configurations. However, in the presence of an external magnetic field, the edge spin orientations become parallel and BLG structure would have ferromagnetic (FM) spin configuration along all zigzag edges, similar to monolayer graphene nanoribbons and flakes \cite{Pisani,Lee,Guo} (see fig. 1c). We considered two different spin configurations in BLG junction. Rectangular BLGs with AFM spin configuration along zigzag edge have a zero net magnetic moment according to the Lieb's theorem \cite{Lieb}. Due to external magnetic field applied perpendicular to BLG, it has a FM spin configuration and a net magnetic moment in the channel. Consequently, the degeneracy between up- and down-spin electrons breaks.  According to the prediction of density functional theory, the spin-correlation length limits long-range magnetic order to 1 nm at room temperature\cite{Yazyev}, therefore in our calculation the semi-infinite ZGNRs do not include edge magnetism and all spin-dependent scatterings are induced by the localized magnetic moments in BLG. In other words, the effect of edge magnetic moments on the zigzag-shaped edges in these ZGNR electrodes was intentionally not included in our calculations in order to emphasize on the role of magnetic field in BLG geometry and its magnetic edge-states in generating spin-polarized currents and GMR.

We showed in Fig. 2 the spin-dependent destiny of states and conductance as some functions of energy for the device depicted in Fig. 1 with AFM and FM spin configurations in BLG. By comparing density of states and conductance for AFM an FM cases, it is clear that due to the influence of external magnetic field on the edge-states in the BLG nanoflake, the conductance and density of states of the BLG junction do not vanish at the band center. Therefore, the electronic structure and conductance strongly change with the spin orientation of the zigzag edge-state. Moreover, the density of states and conductance for the two spin sub-bands are non-degenerate in FM case (see Figs. 2b and 2d). Hence, in the FM case, conductance and energy spectra around the Fermi energy indicate a high conductivity and a spin-polarized behavior. Generally, magnetic transport properties of ZGNR/AB-stacked BLG/ZGNR junction are strongly dependent on the magnetization of the zigzag edge-states in BLG.

In FM case, when the electron conduction through the nonmagnetic ZGNR as electrodes arrives at BLG nanoflake, different scatterings occur for majority- and minority-spin electrons. Hence, the transmission coefficients for spin-up and spin-down electrons are different. For more clarification, we plotted spin polarization ($SP(\varepsilon)= (G_{\uparrow}(\varepsilon)-G_{\downarrow}(\varepsilon))/(G_{\uparrow}(\varepsilon)+G_{\downarrow}(\varepsilon))\times 100$) as a function of energy and with $N_x=12$ and four different widths in Fig. 3. Interestingly, a high spin polarization near the Fermi energy and in a large energy region could be obtained for the AB-stacked BLG junction with different widths. It is clear that, the spin polarization depends on the BLG width. For example, in BLG of widths $N_y=4$ and $N_y=6$, the spin polarization through the junction will be different from that of a BLG of widths $N_y=8$ and $N_y=10$. In summary, BLG of widths  $N_y=4$ and  $N_y=6$ shows a perfect spin polarization in a large energy region. As width increases, some non-spin-polarized local density is produced around the Fermi energy due to the effect of non-magnetic edge state of ZGNR as electrodes, and spin polarization values are therefore strongly modulated by increasing width near the Fermi energy.

\begin{figure}
\centerline{\includegraphics[width=.95\linewidth]{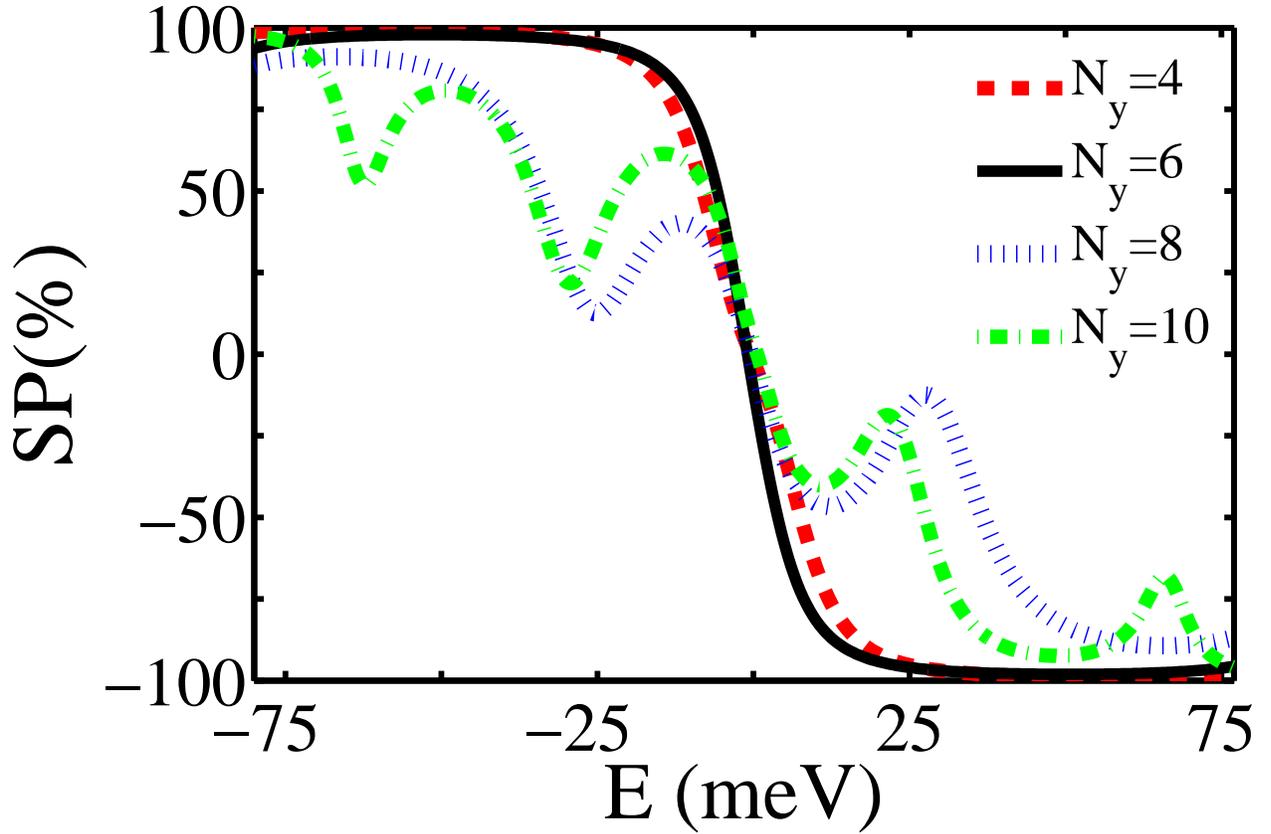}}
\caption{ (Color online) Spin polarization as a function of energy for four BLG nanoflake widths with $N_x=12$ and FM spin configuration.}
\end{figure}

In order to see the sensitivity of the device to a switching magnetic field, we plotted the zero-bias magnetoresistance as $MR= (G_{FM}-G_{AFM})/G_{AFM} \times 100$  as a function of BLG length for four widths, as shown in Fig. 4a. The junction, made entirely of carbon, produces a giant magnetoresistance (GMR) up to $10^6 \%$. Fig. 4a shows the zero-bias magnetoresistance changes in a wide range from $10^2 \%$ to $10^6 \%$. Our results showed that the GMR in the junction is sensitive to the width and length of BLG nanoflakes.
Moreover, we can produce a higher magnetoresistance in BLG junction by using a longer BLG nanoflake.  Interestingly, for a BLG longer than 1 nm, GMR increases as channel length increases and its width decreases, and the GMR could be controlled by changing the BLG length and width. Generally, MR in a AB-stacked BLG strongly increases (decreases) with length (width). As length increases, the total magnetic moment in BLG nanoflake increases and this increase produces a larger spin splitting and GMR in these devices. But as width increases, some non-spin-polarized local density is produced around the fermi energy due to the effect of non-magnetic edge state of ZGNR as electrodes, and therefore decreases GMR in our devices.
\begin{figure}
\centerline{\includegraphics[width=0.95\linewidth]{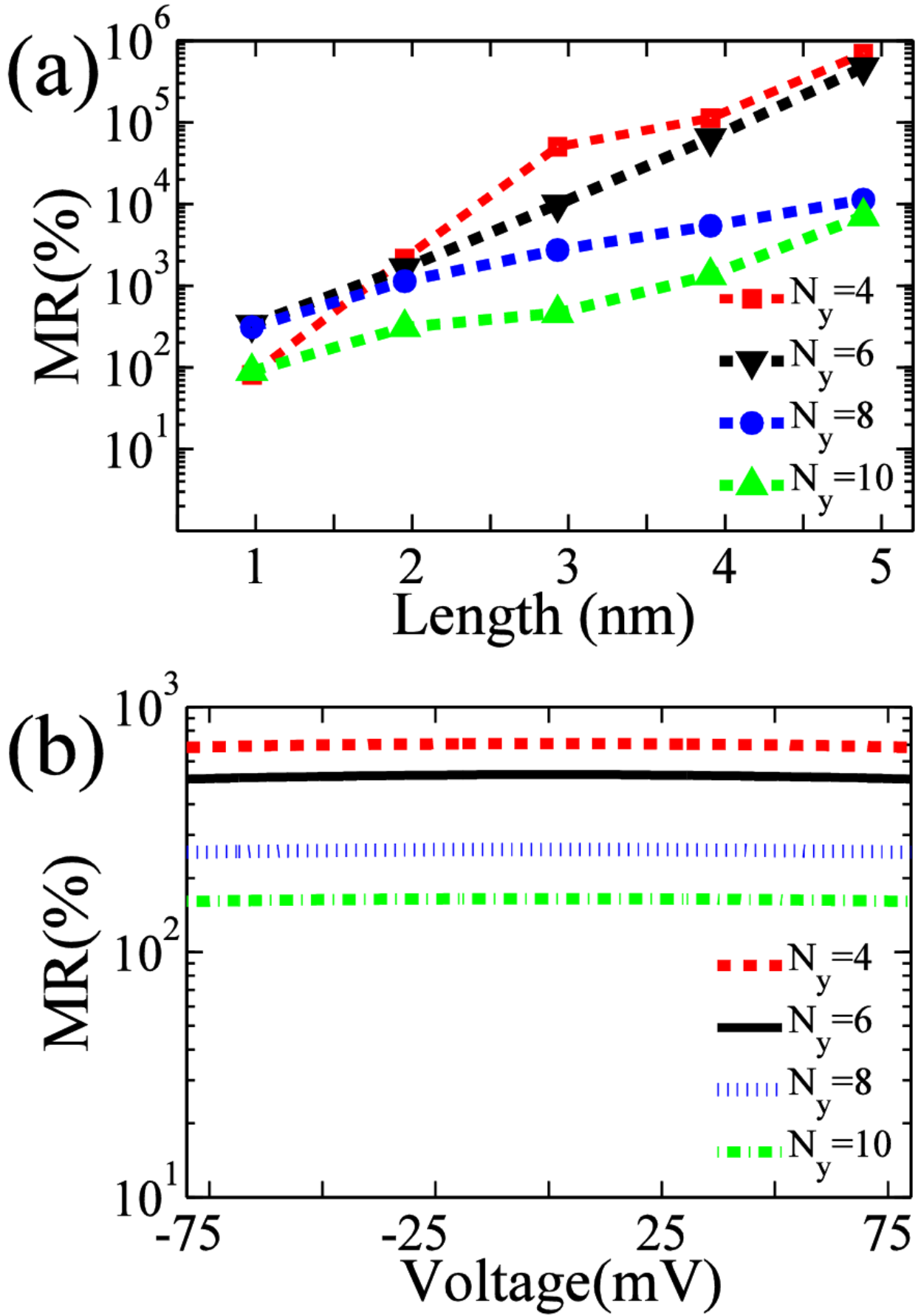}}
\caption{(Color online) (a) The zero-bias magnetoresistance as a function of BLG length for four widths. (b) MR values as a function of bias voltage for four widths and $N_x=12$ at 100 K.
}
\end{figure}

In Fig. 4 (b), we plotted the GMR as a function of bias voltage with $N_x=12$ and various widths at 100 K. For zigzag graphene nanoribbon, the spin correlation length at 100 K is about 3 nm  \cite{Yazyev}. Therefore, spin configuration in a BLG having Nx=12 does not change significantly by temperature. As BLG width increases the conductance value around Fermi energy ($E_F$) in AFM state increases, and at low bias voltage the current also increases, so GMR value decreases from $10^3 \%$to $10^2 \%$ as BLG width increases. Moreover, the GMR slop as a function of bias voltage (Fig. 4b) remains constant by varying the BLG width. Generally, the zigzag edge-state in the AB-stacked BLG is strongly modulated by the applied external magnetic field and produce GMR in a large bias region from -80 to 80 mV. Note that GMR value in our proposed BLG junction is about  that in single-layer ZGNR junction \cite{Kim,Bai}. It suggests that our proposed device might possess some advantages in construction of BLG-based spin devices. Moreover, in order to keep the transport in the ballistic regime, the bias voltage is assumed to be very small, implying that electrons are transmitted through the channel with energies close to the Fermi level $(E_F = 0 eV)$.

The spin-unpolarized zigzag-edge bilayer graphene nanoribbons show a sharp peak at the Fermi energy, and so they have a metallic behavior in AA- and AB-stacking arrangements. However, for the spin resolved solution, AB-stacked and AA-staked cases show different electronic properties. Interestingly, AB-stacked bilayer graphene is semiconducting and metallic depending on the AFM and FM spin configuration, respectively (similar to AB-stacked BLG nanoflake) \cite{Guo}. In contrast to AB-stacked, AA-stacked bilayer graphene in spin-resolved situation is metallic \cite{Guo} and does not show obvious changes in band structure by changing spin configuration. Hence, in the AA-stacked case, the conductance could not be modulated by magnetization of the zigzag edges and this structure could not show a significant magnetoresistance effect.

\section{Conclusion}
The magnetoresistance and coherent spin transport properties of zigzag-edge BLG nanoflake attached between two semi-infinite ZGNR electrodes were investigated by using a tight binding model, a mean field Hubbard Hamiltonian, and the Landauer-Buttiker formalism. Our results showed that by designing a graphene-based structure and utilizing the property of intrinsic magnetic edge-states of BLG, one could generate GMR by applying an external magnetic field in the small part of the junction, without coupling ferromagnetic leads. In AB-stacked BLG, the conductance could be strongly modulated by magnetization of the zigzag-edge BLG and by applying a magnetic field, a perfect spin-polarized current and a GMR up to $10^6 \%$ could be generated. AB-stacked BLG width and length have significant effects on the magnetic transport and GMR properties in ZGNR/AB-stacked BLG/ZGNR. Our results showed that the AB-stacked BLG has a strong modulating ability for the spin-dependent transport and it might possess some advantages in construction of carbon-based spin devices.

\section*{Acknowledgement}
This work is financially supported by university of Kashan under grant No. 468825.

\end{document}